\documentclass[twocolumn,prl,amsmath,amssymb,showpacs,superscriptaddress]{revtex4}
\usepackage{graphicx}
\usepackage{dcolumn}
\usepackage{bm}
\usepackage{latexsym}
\usepackage{amstext}
\usepackage{amsxtra}
\usepackage{color}

\definecolor{darkblue}{rgb}{0, 0, 0.8}
\usepackage[colorlinks=true, breaklinks=true, linkcolor=darkblue, citecolor=darkblue, urlcolor=darkblue]{hyperref}
\newcommand{\doilink}[2]{\href{http://dx.doi.org/#1}{#2}}

\newcommand{\ket}[1]{|#1\rangle}

\begin{document}

\title{Direct measurement of the van der Waals interaction between two Rydberg atoms}

\author{L. B\'eguin}
\affiliation{Laboratoire Charles Fabry, Institut d'Optique, CNRS, Univ Paris Sud,\\
2 avenue Augustin Fresnel, 91127 Palaiseau cedex, France }

\author{A. Vernier}
\affiliation{Laboratoire Charles Fabry, Institut d'Optique, CNRS, Univ Paris Sud,\\
2 avenue Augustin Fresnel, 91127 Palaiseau cedex, France }

\author{R. Chicireanu}
\affiliation{Laboratoire de Physique des Lasers, Atomes et Mol\'ecules, Universit\'e Lille 1, CNRS; \\59655 Villeneuve d'Ascq cedex, France}

\author{T. Lahaye}
\affiliation{Laboratoire Charles Fabry, Institut d'Optique, CNRS, Univ Paris Sud,\\
2 avenue Augustin Fresnel, 91127 Palaiseau cedex, France }

\author{A. Browaeys}
\affiliation{Laboratoire Charles Fabry, Institut d'Optique, CNRS, Univ Paris Sud,\\
2 avenue Augustin Fresnel, 91127 Palaiseau cedex, France }

\date{\today}

\begin{abstract}
We report the direct measurement of the van der Waals interaction between two isolated, single Rydberg atoms separated by a controlled distance of a few micrometers. Working in a regime where the single-atom Rabi frequency for excitation to the Rydberg state is comparable to the interaction, we observe \emph{partial} Rydberg blockade, whereby the time-dependent populations of the various two-atom states exhibit coherent oscillations with several frequencies. Quantitative comparison of the data with a simple model based on the optical Bloch equations allows us to extract the van der Waals energy, and observe its characteristic $C_6/R^6$ dependence. The measured $C_6$ coefficients agree well with \emph{ab-initio} calculations, and we observe their dramatic increase with the principal quantum number $n$ of the Rydberg state.
\end{abstract}


\maketitle

The van der Waals-London interaction $U_{\rm vdW}=C_6/R^6$ between two neutral, polarizable particles separated by a distance $R$ is ubiquitous in nature. It underlies many effects, from the condensation of non-polar gases, to the adhesion of gecko toes~\cite{israelachvili2010}. At a macroscopic scale, measuring thermodynamic quantities of a system gives indirect access to the van der Waals interaction between the constituent atoms or molecules~\cite{hiemenz1997}. Alternatively, one can directly measure the net force between macroscopic bodies resulting from the microscopic van der Waals forces. However, in this case, summing over the underlying $C_6/R^6$ interactions between individual particles results in a potential scaling as $1/L^\alpha$, where $L$ is the separation between the bodies, and $\alpha<6$ a geometry-dependent exponent (e.g. $\alpha=1$ for two spheres with a diameter $D\gg L$)~\cite{israelachvili2010,hiemenz1997}.

At the level of individual particles, spectroscopy of the vibrational levels close to the dissociation limit of a diatomic molecule, analyzed semi-classically, allows to infer the long-range interaction between atoms~\cite{leroy1970}. Progress in atomic physics has made it possible to measure van der Waals interactions between ground-state atoms and a \emph{surface} (scaling as $1/L^3$, or even $1/L^4$ if retardation plays a role) with a variety of techniques~\cite{gsvdw}. However, directly measuring the van der Waals interaction between two ground-state atoms would be extremely challenging, due to their very small interaction. In contrast, Rydberg atoms (atoms with large principal quantum number $n$) exhibit very strong interactions that scale rapidly with $n$. Using this property, the interaction between Rydberg atoms and a surface has been measured at relatively large distances~\cite{anderson1988,sandoghdar1992}. Here, we report on the measurement of the $C_6/R^6$ interaction between two isolated Rydberg atoms prepared in a well defined quantum state.

\begin{figure}
\centering
\includegraphics[width=8.5cm]{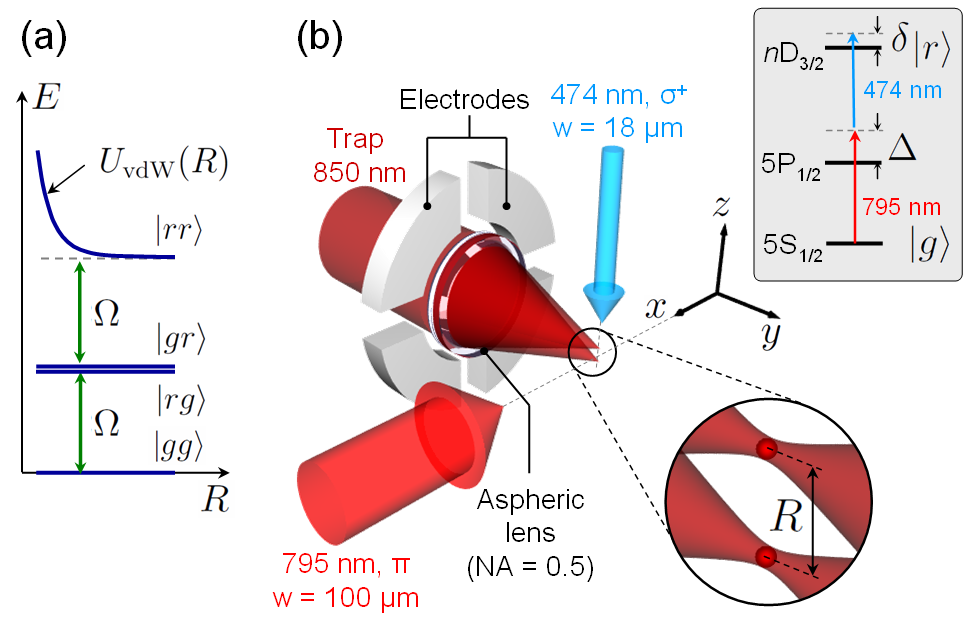}
\caption{(color online). (a): Principle of the Rydberg blockade. The single-atom Rabi frequency is $\Omega$. (b): Experimental setup. Two single atoms are trapped in microscopic optical traps separated by~$R$. Eight electrodes (four of which, facing the ones displayed here, are not shown) provide electric field control. Inset: two-photon excitation scheme (the intermediate-state detuning is $\Delta\simeq2\pi\times740$~MHz).}
\label{fig:fig1}
\end{figure}

The principle of our experiment is the following. We irradiate the pair of atoms with lasers that couple the ground state $\ket{g}$ and the targeted Rydberg state $\ket{r}$ with Rabi frequency $\Omega$. Depending on the relative strength of $U_{\rm vdW}$ and $\hbar\Omega$, two limiting cases can be identified. If $U_{\rm vdW}\ll \hbar\Omega$, the atoms behave independently and the doubly excited state $\ket{rr}$ can be fully populated. On the contrary, when $\hbar \Omega\ll U_{\rm vdW}$, the excitation of  $\ket{rr}$ is off-resonant and thus suppressed (Fig.~\ref{fig:fig1}a), yielding \emph{Rydberg blockade}~\cite{jaksch2000,lukin2001,comparat2010}. This leads to the appearance of ``blockade spheres'' inside of which only one Rydberg excitation can be created. The blockade sphere picture gives an intuitive understanding of non-trivial many-body effects in driven systems. Rydberg blockade has been observed in recent years in extended cold atoms ensembles~\cite{singer2004,heidemann2007,pritchard2010,dudin2012,schauss2012} as well as between two atoms~\cite{urban2009,gaetan2009}.

In the transition region $\hbar\Omega\sim U_{\rm vdW}$, i.e. in the \emph{partial blockade} regime, the blockade sphere picture is too simplistic: the populations of the various many-atom states undergo coherent collective oscillations with several frequencies which depend on $U_{\rm vdW}$. In our two-atom experiment, this allows us to extract the interaction strength. By measuring $U_{\rm vdW}$ for various $R$, we observe its characteristic $1/R^6$ dependence. The measured $C_6$ coefficients, which scale extremely fast with $n$, agree well with \emph{ab-initio} calculations. Our results prove that detailed control over the interactions between Rydberg atoms is possible; this is a prerequisite for applications to high-fidelity quantum information processing~\cite{saffman2010} and quantum simulation using long-range interatomic interactions~\cite{weimar2010}.

We use two single $^{87}{\rm Rb}$ atoms at a temperature of 50~$\mu$K, loaded in 1~mK-deep microscopic optical traps from a magneto-optical trap (Fig.~\ref{fig:fig1}b)~\cite{schlosser2001}. Our new setup is designed to overcome the limitations of the apparatus used in our early studies of Rydberg blockade~\cite{gaetan2009} and entanglement~\cite{wilk2010}. We use an aspheric lens under vacuum~\cite{sortais2007} with numerical aperture 0.5 (focal length 10~mm, working distance 7~mm) to focus two 850~nm trapping beams down to 1.1~$\mu$m ($1/e^2$ radius). The distance $R$ between the traps is varied by changing the incidence angle of the beams on the lens. We calibrate $R$ by measuring the displacement of an image of the trap when changing the incidence of the trapping beams. The resulting uncertainty is below 5\%~\cite{notecalr}.

\begin{figure*}
\centering
\includegraphics[width=15cm]{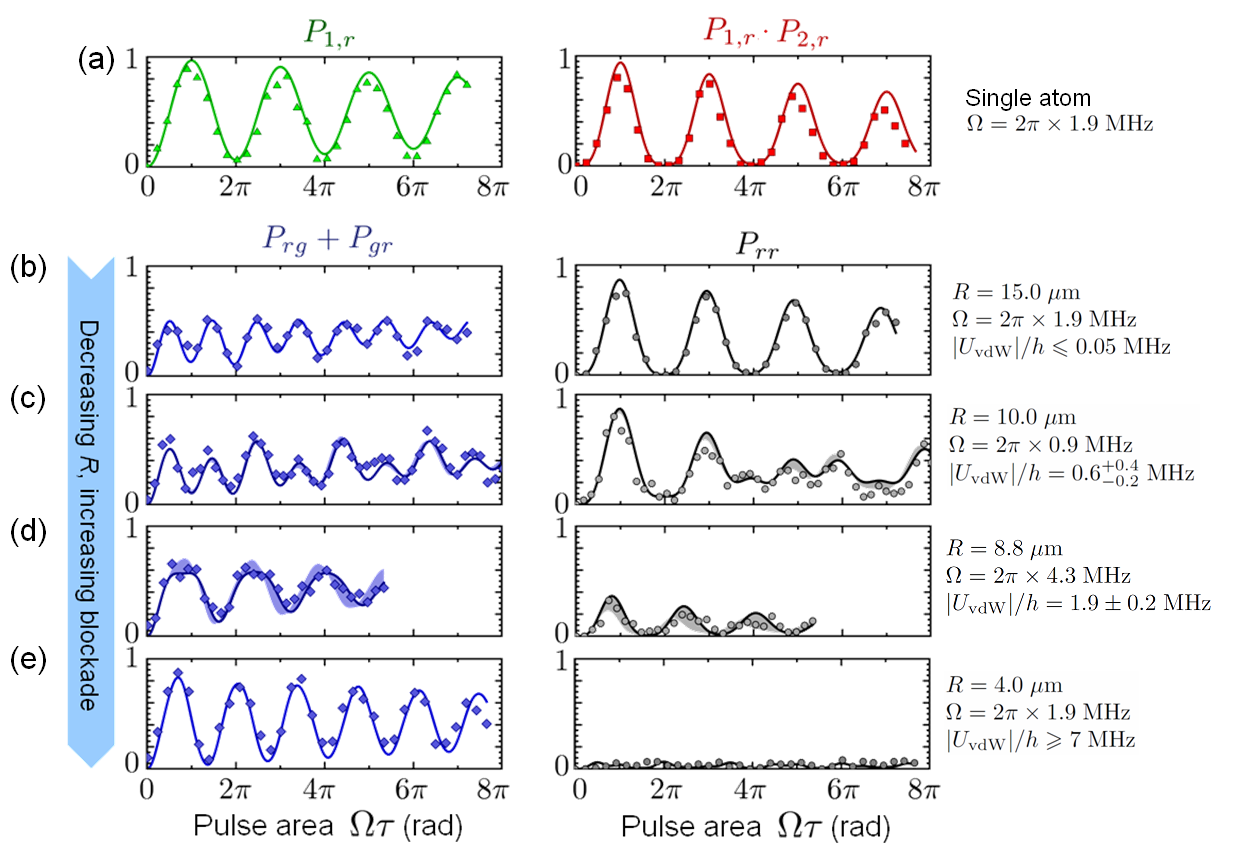}
\caption{(color online). From independent to blockaded atoms, \emph{via} partial blockade, for $n=62$. (a): single-atom Rabi oscillation $P_{1,r}(\Omega\tau)$ between $\ket{g}$ and $\ket{r}$ (green triangles), and product of such excitation probabilities for single atoms in traps 1,2 with the other trap off (red squares). (b)--(e): Probability $P_{rg}+P_{gr}$ to excite \emph{one} atom to the Rydberg state (blue diamonds), and double Rydberg excitation probability $P_{rr}$ (black circles), versus excitation pulse area $\Omega\tau$. From (b) to (e), $R$ decreases, yielding increasing blockade. Solid lines are fits of the data by the solution of the OBEs (see text). Shaded areas correspond to one standard deviation in determining $|U_{\rm vdW}|$ (statistical error bars on $\Omega$, at the $1\%$ level, are not shown).}
\label{fig:fig2}
\end{figure*}

The aspheric lens used to focus the trapping beams also collects the atom fluorescence from each trap on separate photon counters. The two-photon excitation from $\ket{g}=\ket{5S_{1/2},F=2,m_F=2}$ to the Rydberg state $\ket{r}=\ket{nD_{3/2},m_j=3/2}$ (inset of Fig.\ref{fig:fig1}b), described in~\cite{miroshnychenko2010}, yields coherent oscillations with single-atom Rabi frequencies $\Omega/(2\pi)$ in the range 500~kHz--5~MHz (Fig.~\ref{fig:fig2}a). The excitation pulse has a duration $\tau$ (during which the traps are off), and the laser frequencies are adjusted so that the (light-shift corrected) one-atom detuning is $\delta\simeq0$. After excitation, we infer the state of each atom by detecting its presence or absence in its respective trap (atoms in $\ket{r}$ are slightly anti-trapped by the optical potential, which results in their loss). We thus reconstruct the populations $P_{ij}$ of the two-atom states $\ket{ij}$ ($i$, $j$ taking the values $g$, $r$), by repeating each sequence about 100 times~\cite{miroshnychenko2010}.

Our setup was designed to minimize residual electric fields detrimental to Rydberg state manipulations: (i) the aspheric lens surface facing the atoms is coated with a conductive $200$~nm-thick Indium-Tin Oxide layer; (ii) eight independent electrodes allow us to apply electric fields along any direction~\cite{loew2012}. Using Stark spectroscopy on the $\ket{62D_{3/2},m_j=3/2}$ state, we determine that with all electrodes grounded, a residual field of $\sim 150\;{\rm mV/cm}$ (pointing essentially along $x$) was present. After applying appropriate correction voltages, the residual field is below $\sim 5\;{\rm mV/cm}$. This cancellation is critical to the success of the experiment: small (transverse) stray fields mix $\ket{r}$ with other Rydberg states not coupled to light, or exhibiting F\"orster zeros~\cite{walker2008}, thus degrading the blockade.

Figure~\ref{fig:fig2}(b--e) shows the probabilities $P_{rg}+P_{gr}$ to excite only one atom, and $P_{rr}$ to excite both, versus the area $\Omega\tau$ of the excitation pulse, for various $R$ and $\Omega$, in the case $n=62$. In~(b), the atoms are far apart ($R\simeq15\;\mu{\rm m}$) and thus almost independent. Indeed, the probability $P_{rr}$ of exciting both atoms is nearly equal to the product $P_{1,r}\cdot P_{2,r}$ (row (a)), where $P_{i,r}$ is the probability for atom $i$ to be in $\ket{r}$, obtained in a control experiment with only trap $i$. In this case $P_{rg}+P_{gr}$ is expected to oscillate at frequency $2\Omega$ between $0$ and $1/2$, close to what we observe. On the contrary, in~(e), the atoms are close ($R\simeq4.0\;{\rm \mu m}$) and $\Omega\simeq2\pi\times1.9\;{\rm MHz}$ is small enough for almost perfect blockade to occur: at all times, $P_{rr}$ is negligible ($P_{rr}<0.06$, with an average of 0.036), and thus differs substantially from the product of the single-atom excitation probabilities expected for independent atoms. At the same time, $P_{rg}+P_{gr}$ oscillates at $(1.49\pm0.07)\Omega$, close to the expected collective frequency $\sqrt{2}\Omega$~\cite{gaetan2009}.

\begin{figure}[t]
\centering
\includegraphics[width=85mm]{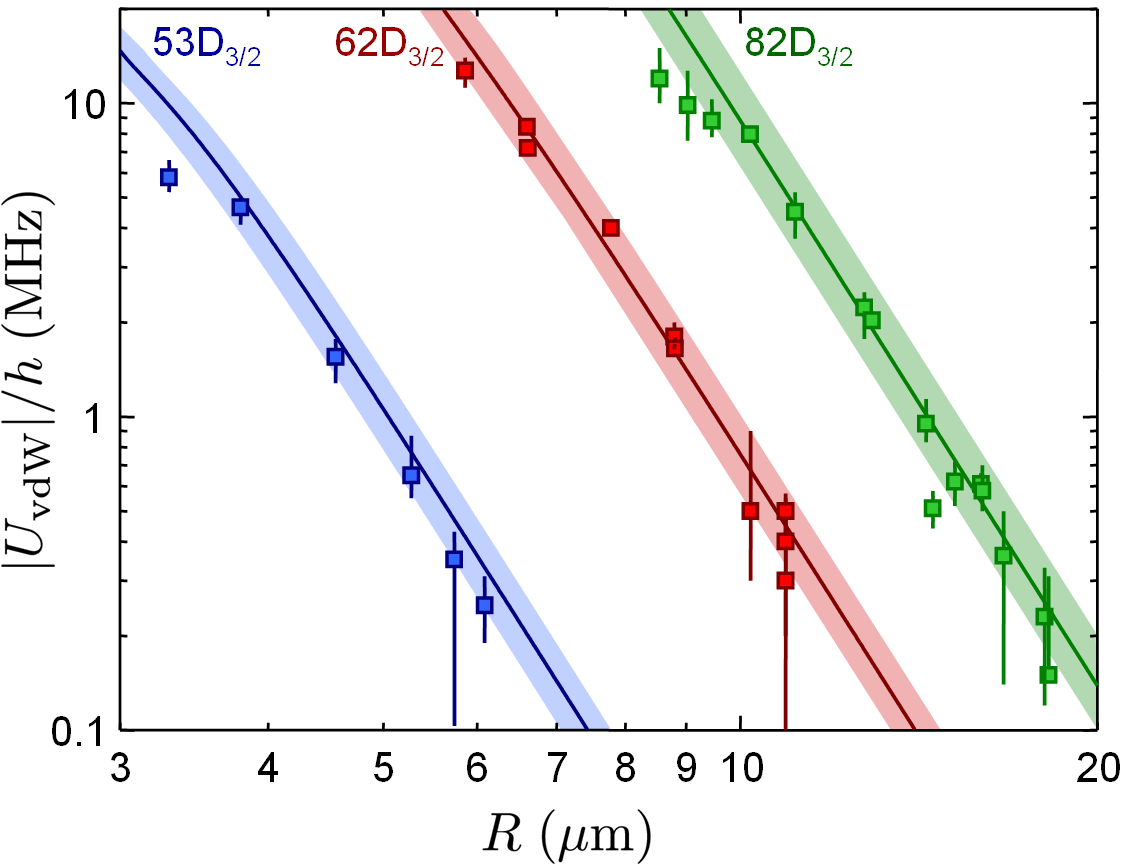}
\caption{(color online). Magnitude $|U_{\rm vdW}|$ of the interaction energy between atoms in $\ket{nD_{3/2},m_j=3/2}$, extracted from the fits of Fig.~\ref{fig:fig2}, versus $R$, for $n=53$ (blue), $n=62$ (red), and $n=82$ (green). Error bars are one standard deviation in the fitted $|U_{\rm vdW}|$. Solid lines: \emph{ab-initio} theoretical expectations (see text) without any adjustable parameter; shaded areas: 5\% systematic uncertainty in~$R$.}
\label{fig:fig3}
\end{figure}

In between those regimes (c and d), blockade is only partial: $P_{rg}+P_{gr}$ and $P_{rr}$ show a more complex behavior, revealing non-trivial two-atom states. We model this dynamics by solving the optical Bloch equations (OBEs)~\cite{ates2007,zhang2012}. Each atom $i$ $(i=1,2)$ is considered as a two-level atom with ground state $\ket{g}_i$ and $nD_{3/2}$ Rydberg state $\ket{r}_i$, coupled by a near-resonant laser with Rabi frequency $\Omega$ (in practice, for our data, the atoms experience, within 5\%, the same $\Omega$). In the basis $\big\{\ket{gg},\ket{gr},\ket{rg},\ket{rr}\big\}$, and using the rotating wave approximation, the hamiltonian is
\begin{equation}
H=
\left(
  \begin{array}{cccc}
    0 & \hbar\Omega/2 & \hbar\Omega/2 & 0 \\
    \hbar\Omega/2 & -\hbar\delta & 0 & \hbar\Omega/2 \\
    \hbar\Omega/2 & 0 & -\hbar\delta & \hbar\Omega/2 \\
    0 & \hbar\Omega/2 & \hbar\Omega/2 & U_{\rm vdW}-2\hbar\delta \\
  \end{array}
\right).
\end{equation}
We omit diagonal terms corresponding to interaction between ground-state atoms (for $R=1\;\mu{\rm m}$ the van der Waals interaction is in the $10^{-8}$~Hz range, the magnetic dipole-dipole interaction in the mHz range) or between a ground-state and a Rydberg atom (in the $1\;$Hz range at $R=1\;\mu {\rm m}$), as they are negligible with respect to the relevant energy scales of the problem.  Experimentally, we observe (especially for large $\Omega$) a small damping of the oscillations, essentially due to off-resonant spontaneous emission via the $5P_{1/2}$ intermediate state. To take this into account, we solve the OBEs for the two-atom density matrix $\dot{\rho}=-i[H,\rho]/\hbar+ \mathcal{L}$. We write the dissipative term as $\mathcal{L}=\mathcal{L}_1\otimes\rho_2+\rho_1\otimes\mathcal{L}_2$, where
\begin{equation}
\mathcal{L}_i=\gamma
\left(
  \begin{array}{cc}
    \rho_{rr} & -\rho_{gr}/2 \\
    -\rho_{rg}/2 & -\rho_{rr} \\
  \end{array}
\right)_i
\end{equation}
is the dissipator for atom $i$ (neglecting dephasing), expressed in the basis $\{\ket{g}_i,\ket{r}_i\}$, and $\rho_i$ the reduced density matrix of atom~$i$. This phenomenological way to include dissipation is sufficient for the present data; a more exact way would include several levels (including the $5P_{1/2}$ state) for each atom, as done in~\cite{miroshnychenko2010,zhang2012}. We neglect cooperative effects such as super-radiance (this is legitimate as $R$ is much larger than the wavelength $\lambda\simeq795\;{\rm nm}$ of the 5S-5P transition which dominates the dissipation via spontaneous emission)~\cite{ates2007}.

The parameters $\Omega$ and $\gamma$ appearing in the model are obtained by fitting single-atom Rabi oscillation data (Fig.~\ref{fig:fig2}a, triangles). The only remaining parameters in the model are $U_{\rm vdW}$ and $\delta$. We treat $U_{\rm vdW}$ as an adjustable parameter to fit the solution of the OBEs to $P_{rg}(t)+P_{gr}(t)$ and $P_{rr}(t)$. Examples of such fits are presented as solid lines in Fig.\ref{fig:fig2}(b-e) (shaded areas show the confidence interval in $U_{\rm vdW}$). We also treat $\delta$ as a free parameter, to account for slow drifts of the lasers~\cite{footnotedelta}. With this method, we obtain only $|U_{\rm vdW}|$, as for $\delta\simeq0$ the sign of the interaction does not affect the dynamics. We have checked that deliberately setting $\delta=0$ and $\gamma=0$ in the fit (thus reducing our analysis to solving an effectively three-level Schr\"odinger equation), yields values of $|U_{\rm vdW}|$ departing by at most 20\% from those above. We checked that the interaction energy yielded by the fits does not depend on the chosen $\Omega$ by doubling or halving it. We emphasize that the convergence of the fit is optimal when $U_{\rm vdW}\sim \hbar\Omega$. Combined with our range of accessible $\Omega$, this means we can determine values of $|U_{\rm vdW}|/h$ in the range 0.1--10~MHz.

Figure~\ref{fig:fig3} shows $|U_{\rm vdW}|$ extracted from such fits versus $R$, for the Rydberg states $\ket{r}=\ket{nD_{3/2},m_j=3/2}$ with $n=53$, $n=62$ and $n=82$. The data is consistent with a power-law of exponent $-6$. Here, the determination of the exponent is much more direct than for interacting ultracold~\cite{heidemann2007} or thermal~\cite{baluktsian2013} ensembles, in which the random distribution of atoms smears out the interaction $R$-dependence. Our results illustrate the dramatic dependence of $U_{\rm vdW}$ with $n$: for instance, changing $n$ from 53 to 62 at given $R$ yields a 50-fold increase in $U_{\rm vdW}$. Fitting the data by $|U_{\rm vdW}| = |C_6|_{\rm exp.}/R^6$ with $|C_6|_{\rm exp.}$ as an adjustable parameter, we obtain the values of Table~\ref{tab:c6}.

To compare our measurements to the theoretical expressions of $|U_{\rm vdW}|$ (solid lines in Fig.~\ref{fig:fig3}), we diagonalize the interaction Hamiltonian, considering only the leading, electric dipole interaction term. From the quantum defects reported in~\cite{li2003}, we compute radial wavefunctions using the Numerov algorithm~\cite{zimmerman1979}. We restrict the Hamiltonian to a two-atom basis $\ket{n_1l_1j_1m_{j1},n_2l_2j_2m_{j2}}$ comprising only states close in energy (up to $ \sim h\times 5$~GHz) from the $\left| n D_{3/2}, m_j=3/2,n D_{3/2}, m_j=3/2\right\rangle $ state, and satisfying $ \left| n-n_{1,2}\right|\leq 4$. This corresponds to (sparse) matrices up to $\simeq 10^{3}\times 10^{3}$~\cite{footnotematrix}.

At the large distances relevant here, the dipole-dipole interaction simply shifts the two-atom levels by a quantity $C_6/R^6$ that can be obtained from second-order perturbation theory. At shorter distances, mixing between adjacent levels occurs~\cite{walker2008}, altering the $1/R^6$ character of the interaction (this can be seen for $n=53$ when $R<4\;\mu{\rm m}$). We obtain the $|C_6|_{\rm th.}$ coefficients of Table~\ref{tab:c6} by fitting the numerically obtained interactions at distances $15<R<20\;\mu{\rm m}$. Our results are in good agreement with second-order perturbation theory calculations~\cite{reinhard2007}. We get an estimate of the uncertainty in $|C_6|_{\rm th.}$ by adding random, uniformly-distributed relative errors of $ \pm 0.5\%$ to radial matrix elements appearing in the hamiltonian. The relative uncertainty is larger ($\sim 10\%$) for $n=53$, due to cancellations of terms with opposite signs. Taking into account error bars, the agreement between our measurement and the calculated $C_6$ is very good. It appears on Fig.~3 that for the largest values of $U_{\rm vdW}$, our experimental determination systematically lies below the theory. An explanation might be that mechanical effects induced by interactions lead to a modification of the dynamics, as recently suggested~\cite{li2013}. Our present analysis neglects these effects and may lead to an underestimation of the actual interaction. Including these effects is left to future work.

In summary, we have directly measured the van der Waals interaction between two isolated Rydberg atoms. The level of control demonstrated here opens exciting perspectives in multi-atom systems, for observing geometry-dependent effects due to the anisotropy of the dipolar interaction~\cite{caroll2004}, or the non-additivity of van der Waals interactions~\cite{cano2012}. It is also a prerequisite for generating high-fidelity, many-atom entanglement using Rydberg blockade, as well as for quantum simulation of long-range interacting spin systems.

We thank A. Guilbaud, P. Roth, F. Moron and F. Nogrette for technical assistance, C. Evellin for contributions at the early stage of this work, Y. R. P. Sortais for invaluable advice in the design of the setup, H. Labuhn and S. Ravets for a careful reading of the manuscript, and J. Vigu\'e for useful discussions. L.~B. is partially supported by the DGA. We acknowledge financial support by the EU (ERC Stg Grant ARENA, AQUTE Integrating project, and Initial Training Network COHERENCE), and by R\'egion \^Ile-de-France (LUMAT and Triangle de la Physique).

\begin{table}[t]
\begin{center}
\begin{tabular}{ccc}
\hline
\hline
$n$ & \qquad $|C_6|_{\rm exp.}\;({\rm GHz\cdot\mu m^6})$ \qquad &  \qquad $|C_6|_{\rm th.}\;({\rm GHz\cdot\mu m^6})$  \qquad\\
\hline
53 & $13.7\pm1.2$ & $16.9\pm1.7$  \\
62 & $730\pm20$ &  $766\pm15$ \\
82 & $8500\pm300$ & $8870\pm150$ \\
\hline
\hline
\end{tabular}
\end{center}
\caption{Experimental and theoretical $|C_6|$ coefficients for $n=53,62$ and $82$.}
\label{tab:c6}
\end{table}


\begin{thebibliography}{30}
\bibitem{notecalr}
The dispersion in the interatomic distances due to the finite temperature has negligible effects. For $T=50\,\mu{\rm K}$, the transverse size in our traps (of radial frequency $\omega_r\sim2\pi\times 100$~kHz) is  $\sim 100$~nm (longitudinal size only contributes to second-order).
\bibitem{footnotedelta}
For our data, the best-fit value of $\delta/(2\pi)$ lies in the range $\pm300$~kHz, compatible with observed frequency drifts of the excitation lasers over the acquisition of a full dataset (a few hours).
\bibitem{footnotematrix}
At the large distances used to extract $C_6$, only a few tens of states contribute significantly. However, to obtain the correct short-distance behaviour, $\sim10^3$ states are necessary.
\end{thebibliography}
\end{document}